# Seismic quiescence and activation prior to the 2025 *M*8.8 Kamchatka, Russia earthquake

**K. Z. Nanjo[1,2,3,4,5], J. Yazbeck[5], I. T. Baughman[5], and J. B. Rundle[5,6,7]**

[1]Global Center for Asian and Regional Research, University of Shizuoka, Shizuoka, Japan

[2]Center for Integrated Research and Education of Natural Hazards, Shizuoka University, Shizuoka, Japan

[3]Institute of Statistical Mathematics, Tokyo, Japan

[4]Japan Agency for Marine-Earth Science and Technology, Yokohama Institute for Earth Sciences, Yokohama, Japan

[5]Department of Physics and Astronomy, University of California, Davis, CA, USA.

[6]Department of Earth and Planetary Sciences, University of California, Davis, CA, USA.

[7]Santa Fe Institute, Fe, NM, USA.

Corresponding author: K. Z. Nanjo (nanjo@u-shizuoka-ken.ac.jp)

ORCID: https://orcid.org/0000-0003-2867-9185 (K.Z.N), https://orcid.org/0000-0003-0223-5260 (J.Y.), 0009-0005-7895-3084 (I.T.B), https://orcid.org/0000-0002-1966-4144 (J.B.R.).

**Key Points:**

- ETAS modeling identified about 20 years of seismic quiescence before the 2025 *M*8.8 Kamchatka mainshock
- A sharp activation began about 10 days before the mainshock, coincident with an *M*7.4 foreshock
- Similar patterns before the 1997 *M*7.8 and 2006 *M*8.3 events suggest a possible recurring preparatory process, though based on few events

## Abstract

The 29 July 2025 Kamchatka earthquake, of magnitude *M*8.8 in the Kamchatka–Kuril subduction system, provides a unique opportunity to investigate the preparatory processes of a great subduction event. Despite Kamchatka's high seismic activity, the long-term evolution of seismicity preceding major ruptures has been poorly documented. Identifying temporal patterns—such as multiyear quiescence and short-term activation—is essential for understanding megathrust failure processes. We applied the Epidemic-Type Aftershock Sequence (ETAS) model and change-point analysis to earthquakes with $M \geq 5$ within a 100-km radius of the 2025 mainshock epicenter, using the Advanced National Seismic System (ANSS) catalog spanning 1975–2025. This approach quantified temporal variations in the seismicity rate and detected statistically significant change points. We found a pronounced approximately 20-year quiescent interval beginning around mid-2003, followed by an abrupt activation that commenced with the *M*7.4 foreshock on 20 July 2025. Similar quiescence-to-activation sequences near the eventual hypocenter have been reported in previous studies for the 1997 *M*7.8 Kronotsky and 2006 *M*8.3 Simushirskoe earthquakes, which ruptured segments immediately north and south of the 2025 rupture, respectively. Together with these earlier studies, our findings suggest that such quiescence-to-activation transitions may represent a recurring preparatory process for megathrust earthquakes in the Kamchatka–Kuril subduction system, although this inference is based on a small number of events. Decadal-scale quiescence has also been reported off Hokkaido, Japan, overlapping the area assessed as having an elevated likelihood of a 17th-century-type *M*9-class earthquake. Given that Hokkaido is in the Kamchatka–Kuril subduction system, tracking a possible transition from quiescence to activation is important for situational awareness and time-dependent hazard assessment.

## Plain Language Summary

On July 29, 2025, a magnitude 8.8 earthquake struck Russia's Kamchatka Peninsula. We examined nearly 50 years of earthquake records to see how activity changed leading up to this event. We found a long quiet spell—about 20 years starting around 2003—followed by a sharp increase in earthquakes about 10 days before the mainshock, including a magnitude 7.4 foreshock. Similar “quiet-then-active” pattern also appeared before nearby major earthquakes in 1997 and 2006. These repeating patterns suggest that in the Kamchatka-Kuril subduction zone, a transition from quiescence to activation may sometimes occur before great megathrust earthquakes. This pattern does not allow exact prediction, but monitoring such changes may help improve situational awareness and time-dependent earthquake hazard assessment. A comparable multi-year quiescence has also been reported off Hokkaido, Japan, in an area considered to have elevated potential for a future *M*9-class earthquake. Because Hokkaido lies in the same subduction system, continued monitoring for changes in seismic activity remains important.

## 1 Introduction

Large subduction earthquakes around the Pacific Rim have repeatedly generated devastating tsunamis and significant crustal deformation, providing key case studies for understanding the preparatory processes of megathrust events. The Kamchatka Peninsula, at the northwestern edge of the Pacific subduction zone, is among the most seismically active regions on Earth, where the Pacific Plate subducts beneath the North American Plate at about 8 cm $yr^{-1}$ (USGS, 2025), where USGS represents the United States Geological Survey. Historical records show multiple events with magnitudes $M\geq8$, including the 1952 Kamchatka *M*9.0 earthquake that produced a trans-Pacific tsunami (Fig. 1a). Despite this long history, the temporal evolution of seismicity preceding such large events remains poorly documented.

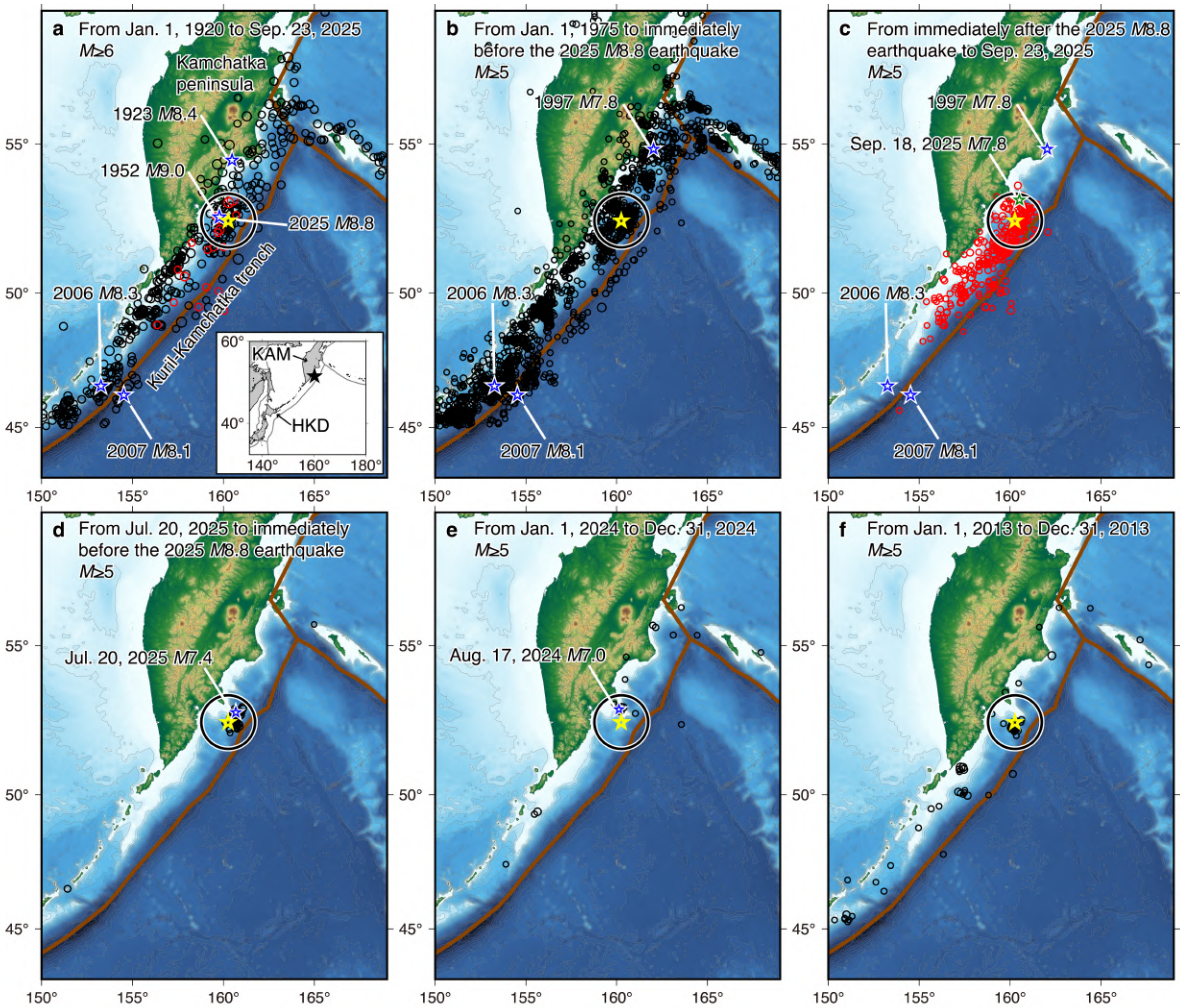

**Figure 1.** Spatial distribution of earthquakes. (**a**) Earthquakes with *M*≥6 (depth ≤ 100 km) from Jan. 1, 1920 up to, but not including, the 2025 *M*8.8 mainshock (black), and from immediately after the mainshock through Sep. 23, 2025 (red). Plate boundaries from Bird (2003) are shown as red curves. The yellow star marks the 2025 *M*8.8 epicenter. Blue stars mark events of *M*≥8. The black circle (100-km radius centered on the *M*8.8 epicenter) delineates the analysis window used in Figs. 2–10. Inset shows the epicenter (star) of the 2025 *M*8.8 epicenter. KAM and HKD represent Kamchatka peninsula and Hokkaido island, respectively. (**b**) Same as (**a**), but for *M*≥5 earthquakes since Jan. 1, 1975 (black). Blue stars mark events with *M*≥7.8. (**c**) Same as (**b**), but for the period from immediately after the 2025 *M*8.8 mainshock to Sep. 23, 2025. The green star marks the Sep. 18, 2025 *M*7.8 event. (**d**) Same as (**b**), but for Jul. 20, 2025 up to, but not including, the 2025 *M*8.8 mainshock. The blue star marks the July 20, 2025 *M*7.4 foreshock. (**e**)

Same as (**b**), but for Jan. 1–Dec. 31, 2024. The blue star marks the Aug. 17, 2024 *M*7.0 event. (**f**) Same as (**b**), but for Jan. 1–Dec. 31, 2013.

On 29 July 2025, an *M*8.8 earthquake occurred off eastern Kamchatka (Fig. 1). The event produced strong shaking and a moderate tsunami, and drew global attention because it ruptured a segment that had been relatively quiet for more than a decade. To investigate its source processes, we analyzed temporal variations in seismicity within a 100-km radius of the 2025 epicenter. Although this paper will illustrates quiescence followed by foreshock activation before the *M*8.8 earthquake, comparable precursory patterns —"quiet-then-active"— have been reported for other great earthquakes, including the 1997 *M*7.8 Kronotsky, Kamchatka event (Sobolev, 2011, 2020), the 2004 *M*9.0 Sumatra event (e.g., Mignan et al., 2006; Sobolev & Lyubushin, 2007; Imoto & Yamamoto, 2008; Katsumata, 2015; Sobolev, 2020), the 2006 *M*8.3 Simushirskoe, Kuril event (Sobolev, 2011, 2020; Katsumata & Zhuang, 2020), the 2010 *M*6.4 Jiashian, Taiwan earthquake (Wen et al., 2016), and the 2011 *M*9.0 Tohoku event (e.g., Katsumata, 2011, 2017; Ogata, 2011; Kato et al., 2012; Nanjo et al., 2012; Nagao et al., 2014).

The 2025 Kamchatka sequence nonetheless provides new insight owing to its clear spatial confinement and the availability of continuous instrumental data spanning more than five decades. Recognizing such patterns is crucial for advancing our understanding of the physics of the earthquake cycle and for improving probabilistic assessments of megathrust occurrence.

In this study, we present a detailed analysis of the temporal evolution of seismicity prior to the 2025 *M*8.8 Kamchatka earthquake. We quantify the long-term quiescence and examine the subsequent activation following the *M*7.4 foreshock. Our results add to growing evidence that temporal variations in seismicity can illuminate the preparatory stages of great earthquakes.

## 2 Materials and Methods

Large earthquakes are often preceded by a period of seismic quiescence—that is, a pronounced, multi-year reduction in regional seismicity (e.g., Rundle et al., 2011, Katsumata and Nakatani, 2021). Recognized as a potential intermediate-term earthquake precursor (e.g., Scholz, 2019), quiescence has been studied for decades. Early reports (e.g., Inouye, 1965; Utsu, 1968; Mogi, 1969; Ohtake et al., 1977; Kanamori, 1981) generally lacked an objective definition, whereas later work introduced quantitative metrics (e.g., Ogata, 1992; Wiemer & Wyss, 1994; Sobolev & Tyupkin, 1997; Ogata, 1999; Rundle et al., 2000; Chen et al., 2005; Holliday et al., 2005; Nanjo et al, 2006a,b; Katsumata, 2011, 2017; Nagao et al., 2011; Rundle et al., 2011; Sobolev, 2011, 2020; Ogata & Tsuruoka, 2016; Katsumata and Nakatani, 2021; Rundle et al., 2022). The widely used metrices includes ZMAP (Wiemer & Wyss, 1994; Katsumata, 2011, 2017; Katsumata and Nakatani, 2021), ETAS (Epidemic-Type Aftershock Sequence) modeling (Ogata, 1992, 1999, Ogata & Tsuruoka, 2016), the RTL/RTM approach (e.g., Sobolev & Tyupkin, 1997; Sobolev, 2011, 2020; Nagao et al., 2011). Collectively, these advances have established seismic quiescence as an objectively quantifiable anomaly that can precede large earthquakes by years to decades.

Sobolev (2020) proposed that large earthquakes are often preceded by a phase of seismic activation—that is, a marked increase in regional seismicity following a period of quiescence. The duration of activation ranges from several days to many months. Sobolev (2020) defined the interval between the end of quiescence and the mainshock as the foreshock-activation phase. In its final stage, events may occur whose locations coincide with the subsequent mainshock source.

Metrics developed to quantify quiescence can likewise be used to characterize activation. Here we adopt the ETAS model (Ogata, 1999), which explicitly represents triggered seismicity and therefore does not require preprocessing declustering, unlike ZMAP or RTL/RTM. Conclusions can be sensitive to the choice of declustering parameters (Wang et al., 2010), so ETAS offers a more self-consistent framework. Before introducing ETAS, we briefly review the Omori–Utsu (OU) law (Utsu, 1961), as follows.

Aftershock rates commonly follow the OU law (Utsu, 1961), $\lambda_{OU}(t)=k(c+t)^{-p}$, where $\lambda_{OU}(t)$ is the conditional intensity (occurrence rate) at time $t$ for events with $M \geq M_{th}$ (minimum threshold magnitude), and $k>0$, $c>0$, and $p>0$. The case $p=1$ reduces to the original Omori (1894) form and is consistent with rate-and-state friction theory (Dieterich, 1994). Values of $p<1$ and $p>1$ correspond to slower and faster temporal decay, respectively. The ETAS model described below represents an earthquake sequence as the superposition of OU-type decays triggered by each prior event.

In the ETAS framework (Ogata, 1999), seismicity is decomposed into a homogeneous Poisson background and a triggered (clustered) component. Each event with magnitude $M_i$ ($\geq M_{th}$) at time $t_i$ ($<t$) contributes $\nu_i(t)=K_0\exp\{\alpha(M_i-M_{th})\}(t-t_i+c)^{-p}$ so that the total rate at time $t$ given the history $H_t=\{(t_i,M_i): t_i<t\}$ is $\lambda(t|H_t) = \mu + \sum_{t_i<t} \nu_i(t)$, where $\mu>0$ is the background rate. The parameter vector $\theta=(\mu, K_0, \alpha, c, p)$ characterizes the sequence; units are $\mu$ in day$^{-1}$, $c$ in days, $\alpha$ and $p$ dimensionless, and $K_0$ having units day$^{p-1}$. Parameters are estimated by maximum likelihood, and model comparison uses the Akaike Information Criterion (AIC; Akaike, 1974). Because $K_0$ depends on $M$ in the model, it is necessary to assume a magnitude at which a value for $K_0$ needs to be known. Throughout this study, $M$=9.0 was assumed for estimating $K_0$ (Nanjo et al., 2023; 2025a,b; 2026).

Goodness of fit can be visualized by comparing observed and model-based cumulative counts in ordinary time and via time-rescaling diagnostics (Ogata & Tsuruoka, 2016). One can visualize this comparison in two types of graphs: one graph using ordinary time and the other using transformed time. Here, transformed time is converted from ordinary time in such a way that the transformed sequence follows the Poisson process (uniformly distributed occurrence times) with unit intensity (occurrence rate). When a good approximation of observed seismicity is presented by the model, one observes an overlap between the observation rate and model rate in both types of graphs.

We performed change-point analysis using ETAS and AIC (Ogata & Tsuruoka, 2016; Akaike, 1974), following Kumazawa et al. (2010, 2019). We define $\Delta AIC=AIC_{single}-AIC_{2stage}$, where $AIC_{single}$ is AIC for the standard (single-stage) ETAS that uses time-invariant parameters and $AIC_{2stage}$ is the AIC for the two-stage ETAS that allows parameters to differ before and after a change time $T_c$ (both fitted to events with $M \geq M_{th}$; see Nanjo et al., 2023; 2025a,b; 2026). The two-stage ETAS is better fitted to the data than the single-stage ETAS model and candidate change is flagged if $\Delta AIC \geq 2q$ at $T_c$, where $q$ is the degree of freedom associated with scanning $T_c$ (Ogata, 1992; Kumazawa et al., 2010). $q$ depends on sample size (number of earthquakes) (Ogata, 1992; Kumazawa et al., 2010). Given $\Delta AIC \geq 2q$, a 68% confidence interval for $T_c$ is obtained (Kumazawa et al., 2010). To calculate it, we considered the likelihood of the two-stage ETAS model, $\exp\{-(AIC_{2stage}+2q)/2\}$ for different $T_c$-values. The normalization of the likelihoods for all $T_c$-values assigned a probability mass, and the central 68% of this mass provided a change point's confidence interval.

**3 Data**

We used the Advanced National Seismic System (ANSS; USGS, 2017) earthquake catalog (https://earthquake.usgs.gov/earthquakes/search/, downloaded on Sep. 25, 2025), which provides source parameters (e.g., hypocenters, magnitudes, phase picks, and amplitudes). From this catalog, we extracted events of $M \geq 5$ with a complete hypocenter and origin-time information occurring between 1920 and Sep. 23, 2025, within 150-169°E, 43-59°N, and depths of 0–100 km. For Fig. 1a, we set a minimum magnitude of $M$=6 to mitigate temporal completeness issues and ensure homogeneity over the full 1920–2025 span.

According to USGS (2025), the Jul. 29, 2025 $M$8.8 event was the latest in a sequence offshore the Kamchatka Peninsula that began 10 days earlier and included an $M$7.4 earthquake on Jul. 20, 2025. As of Jul. 30, 2025, 24 aftershocks of $M \geq 5$ had been recorded, including $M$6.9 and $M$6.3 events (USGS, 2025).

The magnitude–time ($M$–$t$) plots in Fig. 2 confirm this sequence (USGS, 2025). Within a circle of radius $r$=100 km centered on the eventual epicenter, we identified 48 earthquakes of $M \geq 5$ (depth 0–100 km) between the Jul. 20, 2025 $M$7.4 event and immediately before the Jul. 29, 2025 $M$8.8 event (Fig. 1d). After the $M$8.8 mainshock, 150 earthquakes of $M \geq 5$ occurred, including an $M$7.8 event on Sep. 18, 2025 (Fig. 1c). Prior to the Jul. 20, 2025 $M$7.4 event, notable activity since 1975 includes the Aug. 17, 2024 *M*7.0 event (Fig. 1e) and an earthquake swarm with *M* up to 6.1 during May 19–21, 2013 (Fig. 1f).

We chose $r$=100 km because it is the smallest (most compact) area centered on the *M*8.8 epicenter that contains all major seismicity relevant to this study—namely the 1952 *M*9.0 earthquake, the 2013 swarm, the 2024 *M*7.0 earthquake, the Jul. 20, 2025 *M*7.4 event and the subsequent 48 $M \geq 5$ events, and the Sep. 18, 2025 *M*7.8 earthquake (Fig. 1). To verify the sensitivity to the choice of $r$, we also constructed the $M$–$t$ plots for slightly different radii $r$=80 and 120 km (Fig. 2). The same major seismicity features are observed for all three radii, indicating the justification for the choice of $r$=100 km. Therefore, we use $r$=100 km (depth 0–100 km) in the following analyses. However, we examine if the chatacteristics claimed in this paper weaken or disappear beyond $r$=100 km. To examine it, we take different values for $r$ and different study areas in our study.

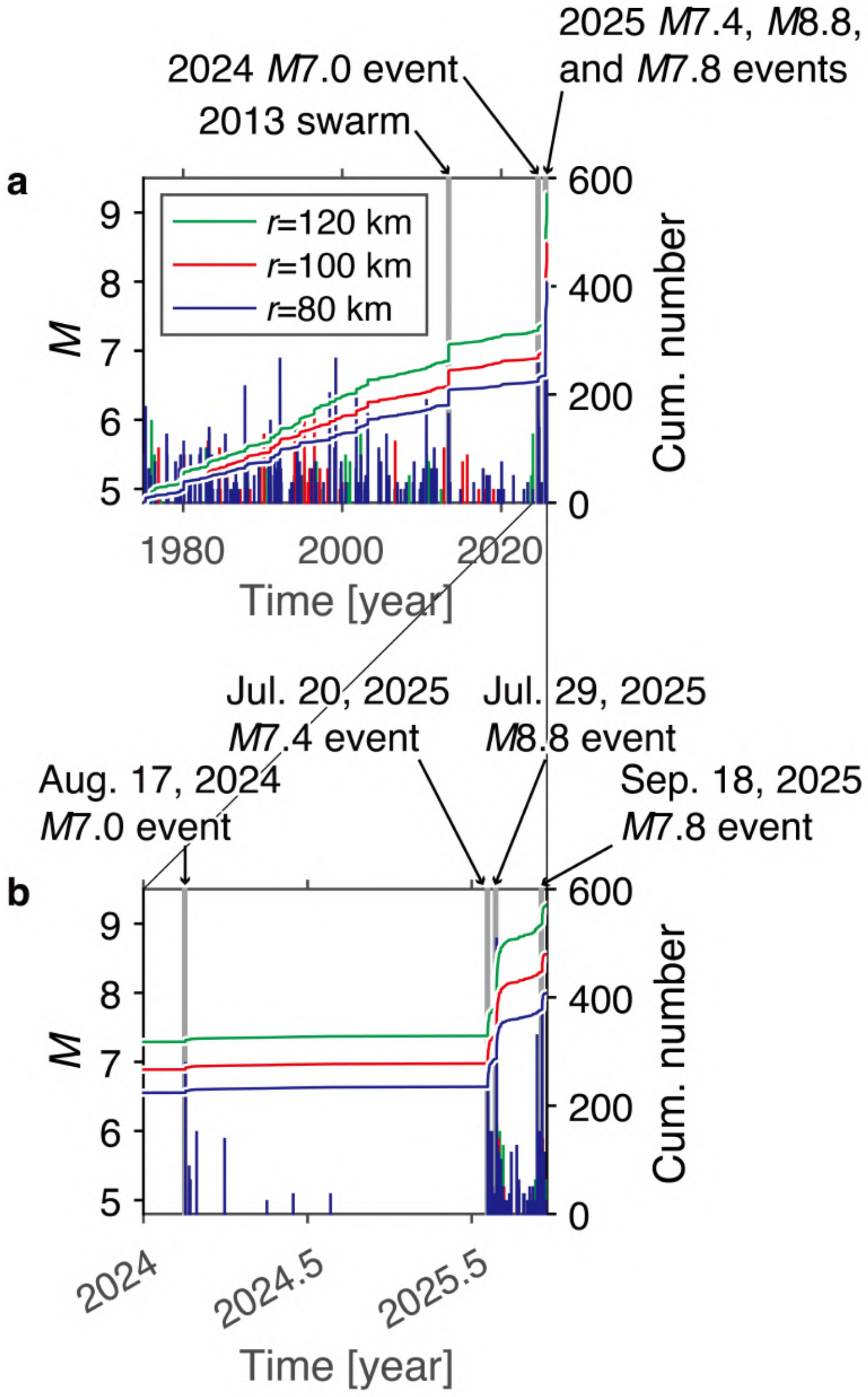

**Figure 2.** Magnitude–time ($M$–$t$) plots for earthquakes with $M \geq 5$ within circles centered on the eventual 2025 *M*8.8 epicenter (depth 0–100 km). Colored bars indicate earthquake magnitudes, and colored curves show the cumulative number of events for radii $r$=80 km (blue), 100 km (red), and 120 km (green). Grey vertical lines mark the May 19–21, 2013 swarm and the Aug. 17, 2024 (*M*7.0), Jul. 20, 2025 (*M*7.4), Jul. 29, 2025 (*M*8.8), and Sep. 18, 2025 (*M*7.8) earthquakes. (**a**) Jan. 1, 1975–Sep. 23, 2025. (**b**) Jan. 1, 2024–Sep. 23, 2025.

The completeness magnitude, $M_c$, above which essentially all earthquakes are detected by a seismic network (e.g., Nanjo et al., 2010), is essential for seismicity-based studies such as the present one. We estimated $M_c$ using a commonly used method based on the Gutenberg–Richter frequency–magnitude relation (Gutenberg & Richter, 1944): given as $\log_{10}N=a-bM$, where $N$ is the cumulative number of earthquakes with a magnitude larger than or equal to $M$, $a$ and $b$ are constants. The magnitude at the deviation point is defined as $M_c$. Below $M_c$, a fraction of events is not detected by the seismic network.

Because our main target study area is confined to the circle of radius $r$= 100 km centered on the eventual epicenter of the 2025 $M$8.8 earthquake (depths 0–100 km), we assessed $M_c$ within this region for three representative periods: Jan. 1, 1920–Dec. 31, 1959; Jan. 1, 1975–Dec. 31, 1989; and Jan. 1, 2020–Dec. 31, 2024. The corresponding frequency–magnitude distributions (Fig. 3) yield $M_c$=5.9, 4.9, and 4.2, respectively. This monotonic decrease of $M_c$ reflects the improving detection capability of the relevant seismic networks over time. Because $M_c$ decreases monotonically with time, its highest value within any span occurs in the earliest sub-period; a magnitude threshold that is complete there is therefore complete throughout the span. The 1920–1959 value ($M_c$=5.9) supports the $M$≥6 threshold adopted for the map in Fig. 1a back to 1920. The 1975–1989 value ($M_c$=4.9) confirms that the $M$≥5 catalog used in the analysis is complete throughout the 1975–2025 study period.

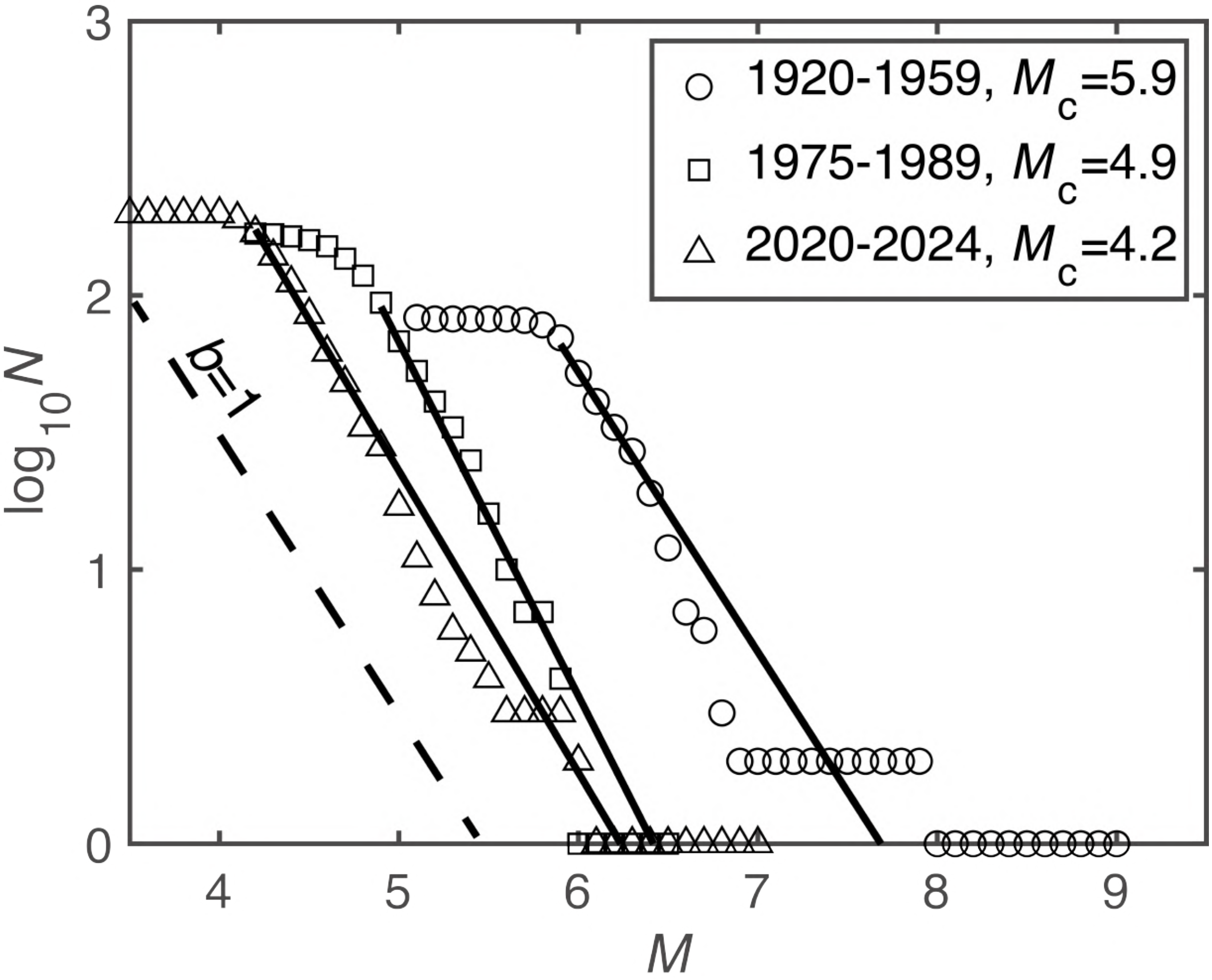

**Figure 3.** log*N* as a function of *M*. Earthquakes in the circle of radius *r*= 100 km centered on the eventual epicenter of the 2025 *M*8.8 earthquake (depths 0–100 km) were used. We considered three representative periods: Jan. 1, 1920–Dec. 31, 1959, Jan. 1, 1975–Dec. 31, 1989, and Jan. 1, 2020–Dec. 31, 2024. We show the GR fits to the respective earthquake samples with *a*=7.84, *b*=1.02±0.10, and $M_c$=5.9 (circle), with *a*=8.28, *b*=1.29±0.10, and $M_c$=4.9 (square), and with *a*=6.86, *b*=1.10±0.08, and $M_c$=5.9 (triangle) are shown. Also included for a reference is the straight line of the slope of 1 (*b*=1).

## 4 Results

Figure 4 shows ETAS fits to earthquakes with $M \geq 5$ ($M_{th}=5$) from Jan. 1, 1975 to Sep. 23, 2025, within a 100-km radius of the eventual $M$8.8 epicenter. The panels compare the observed occurrence rate (black) with the ETAS rate (red) and reveal a sustained deviation followed by a return to baseline prior to the $M$8.8 mainshock. Following Ogata & Tsuruoka (2016) and Kumazawa et al. (2019), we used two time intervals: a target interval, over which the parameter θ was estimated, and a preceding precursory interval to account for long-memory effects of earlier events implied by the OU kernel. Letting "day 0" denotes 00:00:00 on Jan. 1, 1975, we set the precursory and target intervals to 0–$S$ and $S$–$T$, respectively, with $S$=0.01 days (Jan. 1, 1975 00:14:25) and $T$=18528.72 days (Sep. 23, 2025 17:15:22, the last event in the analyzed earthquakes).

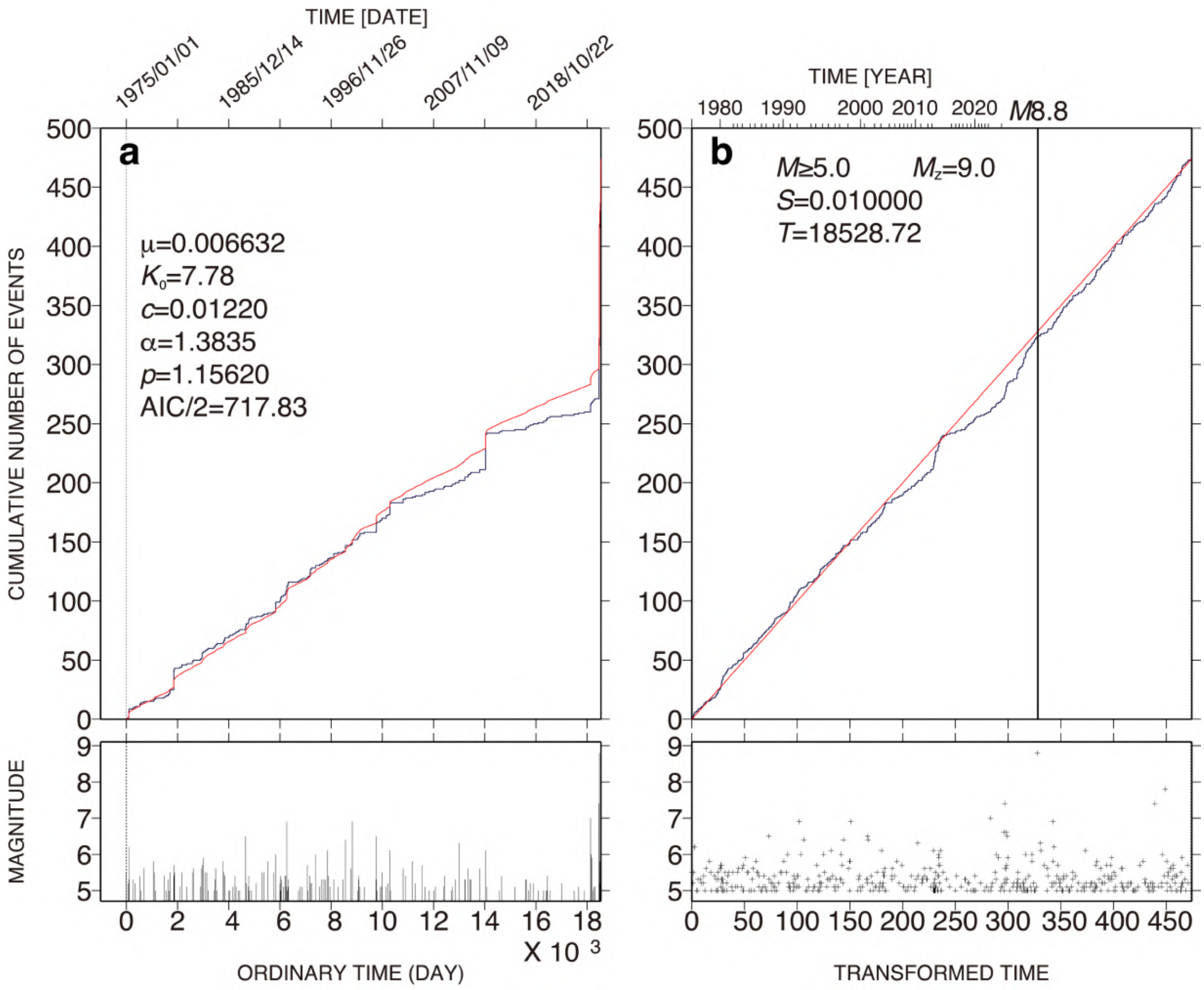

**Figure 4.** ETAS modeling over the full study window. (**a**) Cumulative count of earthquakes with $M$≥5 ($M_{th}$=5) within a 100-km radius of the $M$8.8 epicenter (large black circle in Fig. 1) as a function of ordinary time. The ETAS-based cumulative curve (red) is fitted to the observations (black) over the target interval from Jan. 1, 1975 to Sep. 23, 2025. Day 0 is Jan. 1, 1975 00:00:00, $S$=0.01 days (Jan. 1, 1975 00:14:25, vertical dashed line), and $T$=18,528.72 days (Sep. 23, 2025 17:15:22, the last event in the analyzed earthquakes). The small panel below shows the magnitude–time ($M$–$t$) diagram for the same data. (**b**) Same as (**a**), but plotted in transformed time. See the Materials and Mehods section for explanation of transformed time.

The deviation and subsequent return are more apparent in transformed time (Fig. 4b) than in ordinary time (Fig. 4a). The deviation begins between 10,000 and 11,000 days (May 19, 2002–Feb. 12, 2005), and the return initiates around the $M$7.0 event in 2024 and the $M$7.4 event on July 20, 2025. These features are consistent with a phase of seismic quiescence followed by activation prior to the $M$8.8 mainshock.

To identify the most probable onset of quiescence, we performed change-point analysis (Fig. 5). We scanned candidate change times $T_c$ over Jan. 1, 1990 to Jan. 1, 2010 (in days since Jan. 1, 1975), fitting both the two-stage and single-stage ETAS models to events from Jan. 1, 1975 up to, but not including, the 2025 $M$7.4 event. Instances with $\Delta AIC \geq 2q$ indicate the two-stage model fits better. The 68% confidence interval for the change point (blue band) is $T_c$=10,100–12,100 days (Aug. 27, 2002–Feb. 17, 2008). ΔAIC values were mostly below $2q$ and increasing before $T_c \approx$10,100 days. These values remained above $2q$, and were roughly stable thereafter. A smoothing discontinuity appears near the lower bound of the interval. Considering that the maximum ΔAIC occurs at $T_c$=10,400 days (Jun. 23, 2003; red circle in Fig. 5), we regard $T_c$=10,400 days as the most probable onset of quiescence.

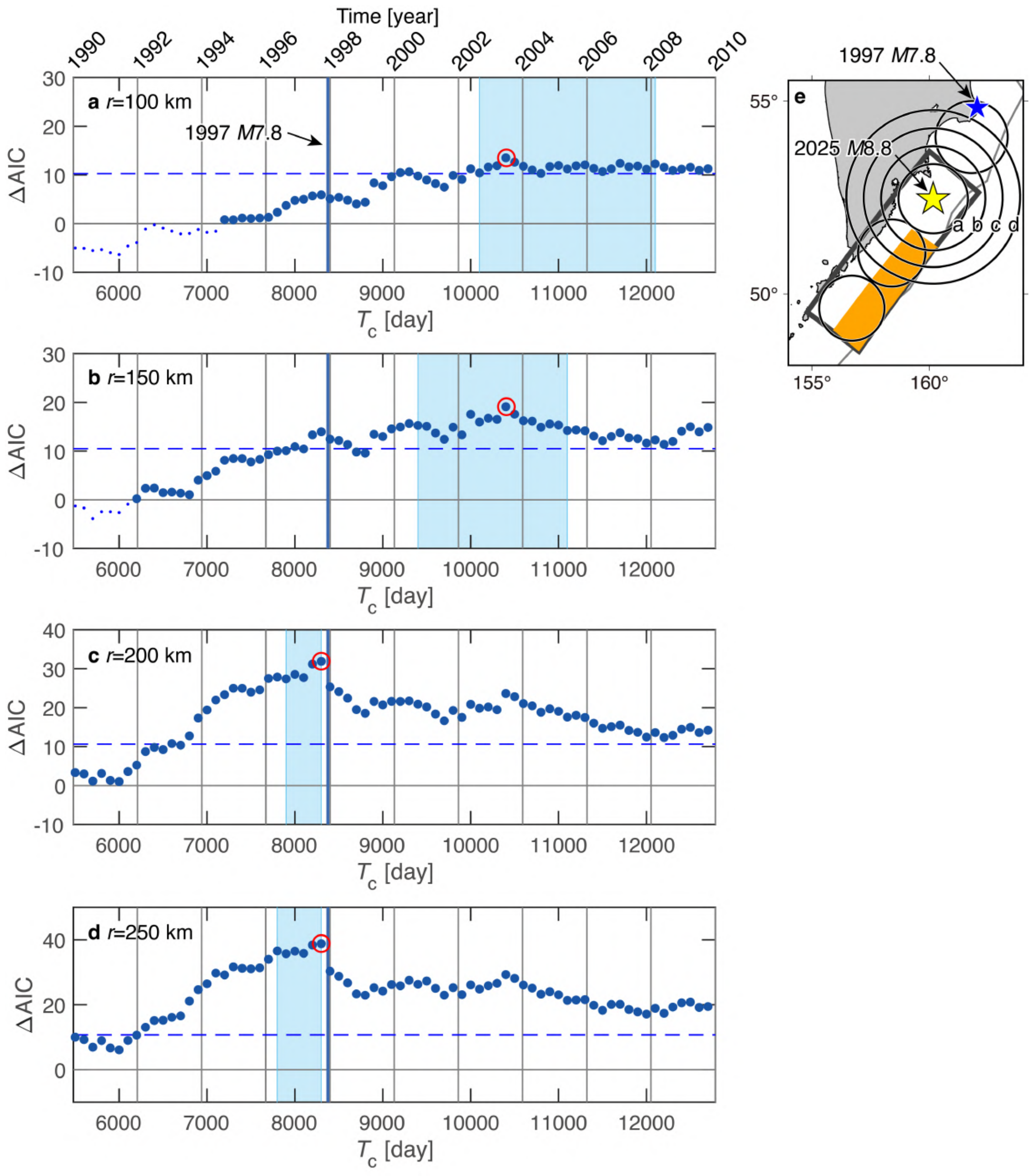

**Figure 5.** ΔAIC as a function of the candidate change time $T_c$ scanned over Jan. 1, 1990–Jan. 1, 2010 (in days since Jan. 1, 1975). The two-stage and standard (single-stage) ETAS models are fitted to events with $M \geq 5$ ($M_{th}=5$) from Jan. 1, 1975 up to, but not including the *M*7.4 foreshock. Thick vertical line indicates the moment of the Dec. 5, 1997 *M*7.8 earthquake (blue star in **e**). Thin vertical ticks indicate Jan. 1 of 1990, 1992, …, 2010. (**a**) To calculate ΔAIC values,

we used seismicity in the circle of $r$=100 km, centered on the eventual epicenter ("a" in **e**). The red circle marks $T_c$=10,400 days. In some cases, the two-stage optimization did not fully converge; when a ΔAIC value was nevertheless returned for that $T_c$, it is plotted as a small dot. The blue band ($T_c$=10,100–12,100 days; Aug. 27, 2002–Feb. 17, 2008) denotes the 68% confidence interval for the change point. The horizontal dashed line shows the 2$q$ threshold. The horizontal solid line indicates ΔAIC=0. (**b**) Same as **a** for $r$=150 km. The red circle marks $T_c$=10,400 days. (**c**) Same as **a** for $r$=200 km. The red circle marks $T_c$=8,300 days. (**d**) Same as **a** for $r$=250 km. The red circle marks $T_c$=8,300 days. (e) Map showing the circles used in **a**-**d**. Yellow and blue stars indicate the 2025 $M$8.8 epicenter and 1997 $M$7.8 epicenter, respectively. Also included for references are the fault model (black rectangle) of the 2025 Kamchatka earthquake (Fujii & Satake, 2026), the high slip (≥4m) region (orange region) of this model, two circles of $r$=100 km within the model.

Figure 6 further corroborates quiescence using three intervals: precursory 0–S=0.01 days (Jan. 1, 1975 00:00:00–00:14:25), target $S$–$T$ (=$T_c$=10,400 days) (through Jun. 23, 2003 00:00:00), and prediction $T$–$T_{end}$ (=18,417.61 days=Jun. 4, 2025 14:39:36; the last event before the $M$7.4 foreshock). We extrapolated the ETAS rate through the prediction interval using θ estimated over the target interval and compared it with observations. The observed cumulative curve generally tracks the lower boundary of the 95% confidence envelope (parabolic bounds). A swarm with $M$ up to 6.1 during May 19–21, 2013 (Fig. 1f) temporarily elevated the rate, but after the swarm ceased, the cumulative curve approached the envelope's lower boundary asymptotically.

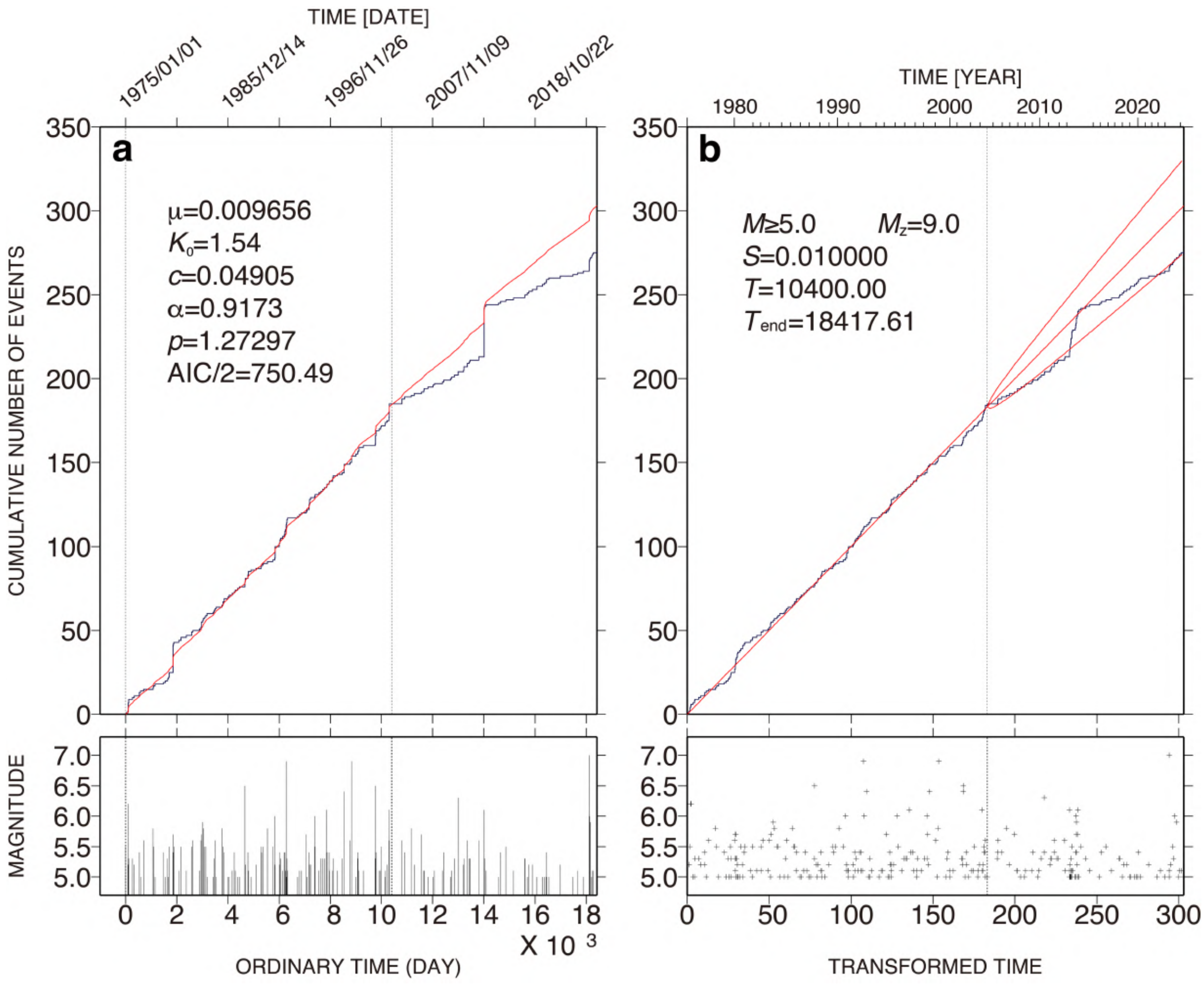

**Figure 6.** Seismic quiescence diagnosed by ETAS modeling for *M*≥5 ($M_{th}$=5) earthquakes within a 100-km radius of the *M*8.8 epicenter. (**a**) Cumulative counts (black) and ETAS fit (red) over the target interval from *S*=0.01 days (Jan. 1, 1975 00:14:25, first vertical dashed line) to *T*=10,400 days (Jun. 23, 2003 00:00:00, second vertical dashed line) plotted in ordinary time, with the ETAS curve extrapolated through the prediction interval ($T$-$T_{end}$) up to immediately before the Jul. 20, 2025 *M*7.4 foreshock, where $T_{end}$=18,417.61 days (the last event before the *M*7.4 foreshock). (**b**) Same as (**a**), but in transformed time; the vertical dashed line marks Jun. 23, 2003. The parabolic envelope denotes the 95% confidence bounds for the ETAS extrapolation.

So far, our analyses have been restricted to *r*=100 km. Whether the quiescence weakens or disappears at larger *r* is itself informative about the preparatory process of a megathrust earthquake, so we repeated the change-point analysis for *r*=150, 200, and 250 km (Fig. 5). At *r*=150 km the maximum ΔAIC remains at $T_c$=10,400 days (Jun. 23, 2003), as for *r*=100 km. At *r*=200 and 250 km, however, the 2003 peak is no longer dominant; the maximum ΔAIC instead falls at $T_c$=8,300 days (Sep. 22, 1997). To locate the origin of this shift, we repeated the analysis for two *r*=100 km circles displaced north and south of the source circle (Fig. 7a,b,e). The epicenter of the Dec. 5, 1997 *M*7.8 Kronotsky earthquake lies within the northern circle (Fig. 7a,e). The northern circle yields a change point at $T_c$=8,800 days (Feb. 4, 1999), whereas the southern circle—displaced toward the region of largest 2025 coseismic slip (Fujii & Satake, 2026)—shows no significant change point (no ΔAIC above 2q). We also repeated the analysis for a *r*=100 km circle displaced further south of the source circle (Fig. 7c), showing again no significant change point.

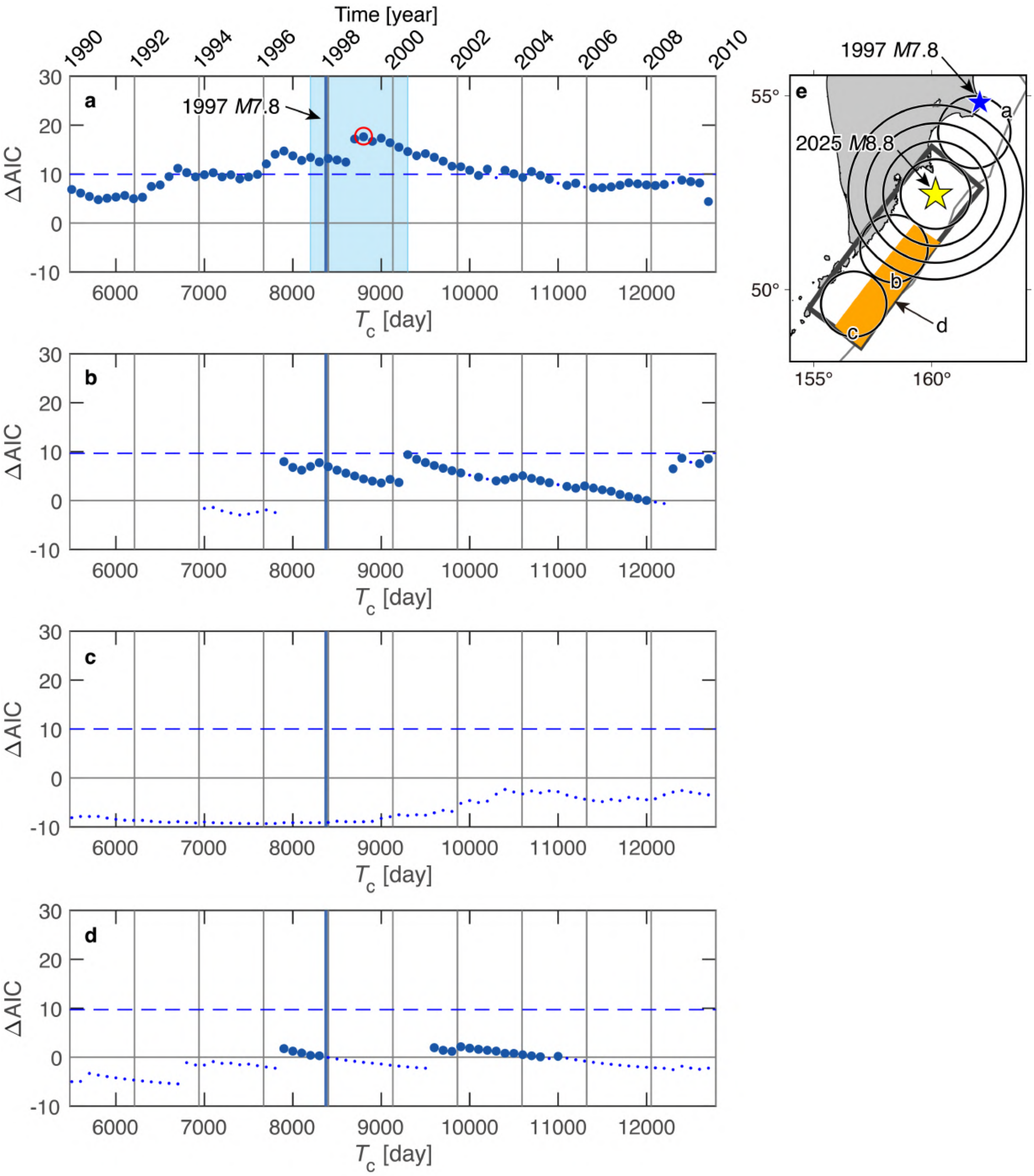

**Figure 7.** Same as Fig. 5 for regions along the plate boundary. (**a**) Result obtained for the circle (center at 161.8°, 54.1°) of $r$=100 km, located north of the circle ($r$=100 km), centered at the 2025 $M$8.8 earthquake (yellow star). (**b**) Same as (**a**) for the circle (center at 156.7°, 49.6°) of $r$=100 km, located south. (**c**) Same as (**c**) for the circle (154.7°, 48.3°), located south of the circle considered for (**c**). (**d**) Same as (a) for the high slip (≥4m) region (orange region in **e**) of the 2025

Kamchatka fault model (black rectangle in **e**) (Fujii & Satake, 2026). (**e**) Same as Fig. 5**e** but the different regions are numbered.

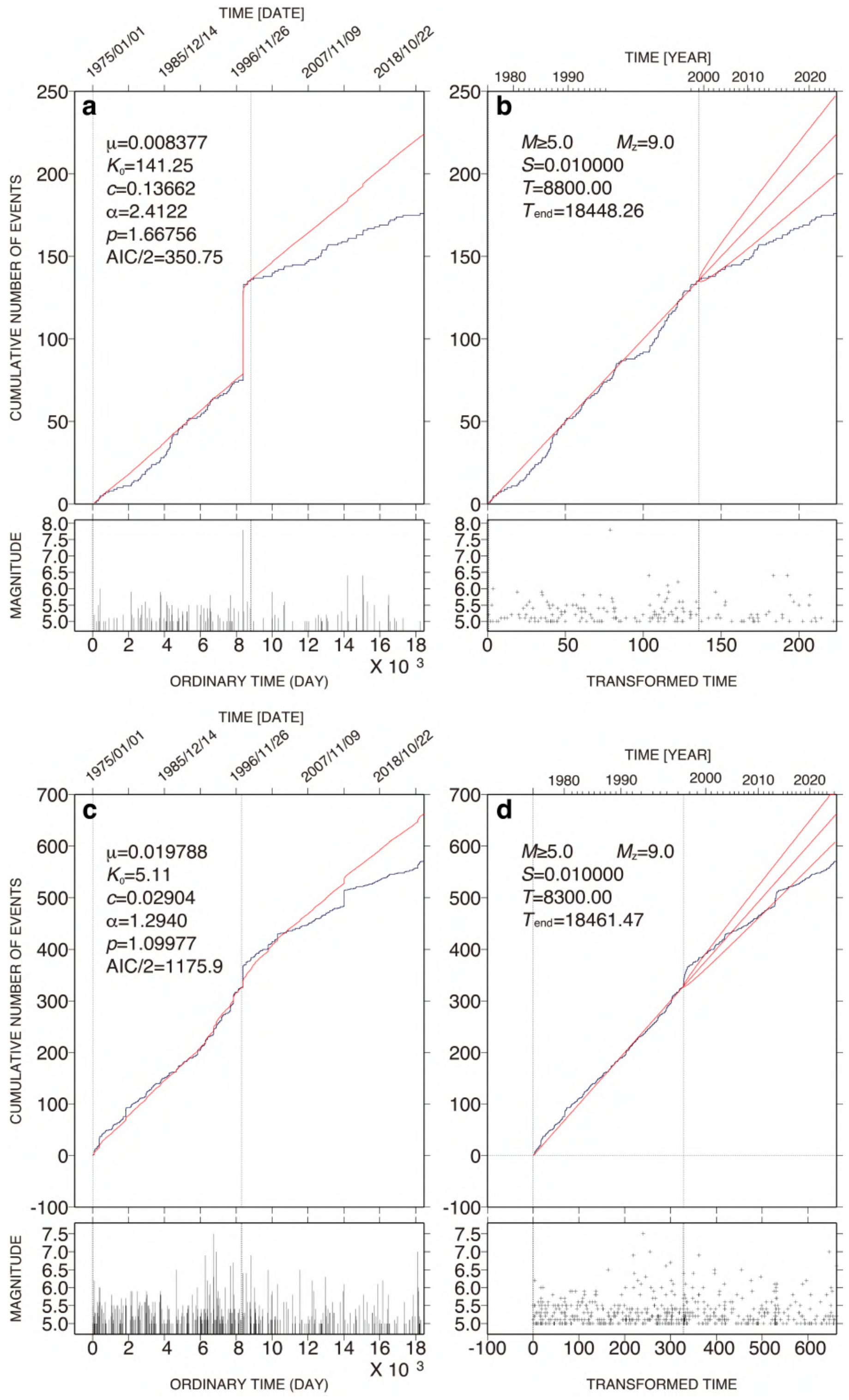

**Figure 8.** Seismic quiescence diagnosed by ETAS modeling for *M*≥5 ($M_{th}$=5) earthquakes within a 100-km radius whose circle is located north to the circle (*r*=100 km), centered at the *M*8.8

epicenter in **a** and **b** and within a 250-km radius of the $M$8.8 epicenter in **c** and **d**. (**a**) Cumulative counts (black) and ETAS fit (red) over the target interval from $S$=0.01 days (Jan. 1, 1975 00:14:25, first vertical dashed line) to $T$=8,800 days (Feb. 4, 1999 00:00:00, second vertical dashed line) plotted in ordinary time, with the ETAS curve extrapolated through the prediction interval ($T$-$T_{end}$) up to immediately before the Jul. 20, 2025 $M$7.4 foreshock, where $T_{end}$=18,448.26 days (the last event before the $M$7.4 foreshock). (**b**) Same as (**a**), but in transformed time; the vertical dashed line marks Feb. 4, 1999. The parabolic envelope denotes the 95% confidence bounds for the ETAS extrapolation. (**c**) The target interval is from $S$=0.01 days (Jan. 1, 1975 00:14:25, first vertical dashed line) to $T$=8,300 days (Sep. 22, 1997 00:00:00, second vertical dashed line). $T_{end}$=18,461.47 days is the timing of the last event before the $M$7.4 foreshock. (**d**) Same as (**c**), but in transformed time; the vertical dashed line marks Sep. 22, 1997.

Figure 8a,b, which is based on data in the circle of $r$=100 km for Fig. 7a, shows seismic quiescence using three intervals: precursory 0–S=0.01 days (Jan. 1, 1975 00:00:00–00:14:25), target $S$–$T$ (=$T_c$=8,800 days) (through Feb. 4, 1999 00:00:00), and prediction $T$–$T_{end}$ (=18,448.26 days=Jul. 5, 2025 6:11:36; the last event before the $M$7.4 foreshock). After $T$ (=$T_c$=8,800 days), seismicity rate was lower than the ETAS rate (black curve is below the red parabola).

Figure 8c,d, which is based on data for $r$=250 km (Fig. 5d), shows seismicity changes using three intervals: precursory 0–S=0.01 days (Jan. 1, 1975 00:00:00–00:14:25), target $S$–$T$ (=$T_c$=8,300 days) (through Sep. 22, 1997 00:00:00), and prediction $T$–$T_{end}$ (=18,461.47 days=Jul. 20, 2025 6:44:37; the last event before the $M$7.4 foreshock). Immediately after $T$=$T_c$=8,300 days, there is a small but sharp increase of seismicity rate (black curve is above the red parabola). After this increase, seismicity rate was comparable to the ETAS rate (black curve is within the red parabola) and finally seismicity rate was lower than the ETAS rate (black curve is below the red parabola).

The change points at larger radii and in the northern circle appear only when the Kronotsky source region is brought into or close to the window—by enlarging $r$ to 200–250 km (Fig. 5c,d) or by displacing the circle northward (Fig. 7a)—and are absent for the $r$≤150 km source circles and the southern circle. They fall at $T_c$=8,300 days (Sep. 22, 1997) for $r$=200–250 km and at $T_c$=8,800 days (Feb. 4, 1999) for the northern circle, bracketing the Kronotsky earthquake itself ($T_c$=8,374 days, Dec. 5, 1997). We therefore attribute them to the seismicity changes accompanying the 1997 Kronotsky earthquake rather than to the 2003 quiescence of the 2025 source region. Conversely, the $T_c$=10,400 days (Jun. 23, 2003) quiescence appears only for circles of $r$≤150 km centered on the eventual epicenter and is absent in the southern circle, identifying it as a localized feature of the immediate source region and linking it to the preparation of the 2025 $M$8.8 earthquake. This dependence on radius and location shows that $r$=100 km is not an a posteriori choice tuned to produce the signal, but the scale at which the localized 2003 quiescence is resolved.

We illustrate the slip distribution of the 2025 $M$8.8 Kamchatka mainshock from Fujii and Satake (2026), digitizing the fault-model area and the high-slip (≥4 m) region (Figs. 5e, 7e). The slip near the epicenter was small, whereas the largest slip occurred 200–400 km to the southwest; the 2025 rupture nucleated near the epicenter and propagated unilaterally about 500 km to the southwest (Fujii & Satake, 2026; Liu et al., 2026). Applying the change-point analysis to the high-slip region (Fig. 7d) yields no significant change point, showing that the 2003 quiescence is confined to the epicentral region and is absent from the region of largest slip.

Because visual inspection of Fig. 4 does not uniquely identify the onset of activation—although times around the 2024 $M$7.0 and July 20, 2025 $M$7.4 events are plausible—we applied the same change-point procedure for activation (Fig. 9). Here we analyzed the period from Jan. 1, 2014 up to, but not including the $M$8.8 mainshock, and found that stable evaluation of $AIC_{2stage}$ for all candidate $T_c$ necessitated reducing the number of free parameters. We therefore fixed α=1.27 (as estimated in Fig. 6), holding it common to both subperiods (independent of $T_c$). Given this predefined α, we optimized (μ, $K_0$, $c$, $p$) separately on each side of $T_c$ to compute $AIC_1$ and $AIC_2$, with $AIC_{2stage}$ (=$AIC_1$+$AIC_2$). For comparison, we also computed $AIC_{single}$ for the single-stage ETAS model under the same constraint (α=1.27) by optimizing (μ, $K_0$, $c$, $p$) over the full interval.

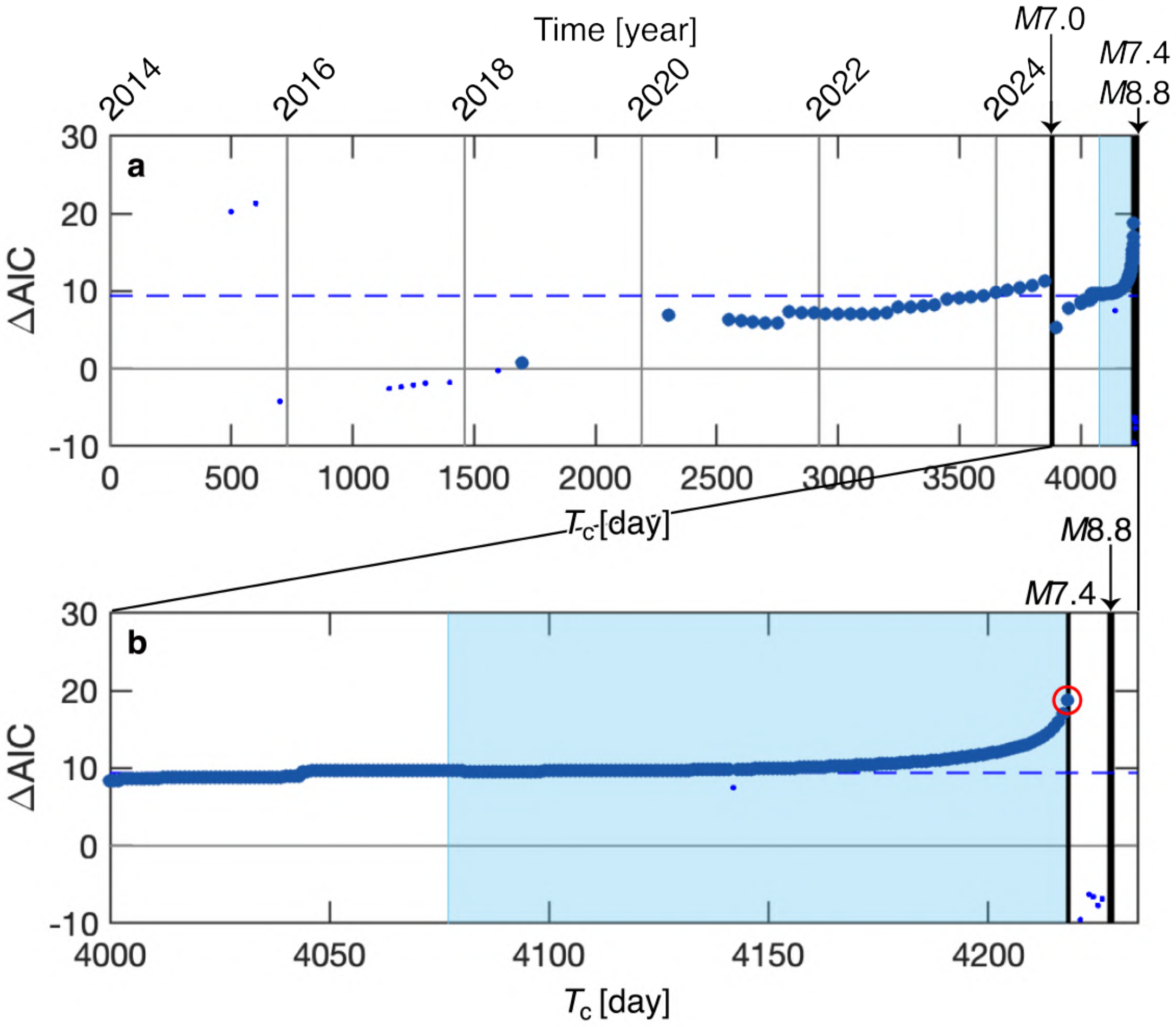

**Figure 9.** Same as Fig. 5a, but for Jan. 1, 2014 to immediately before the *M*8.8 mainshock. The two-stage and standard (single-stage) ETAS models are fitted to events with $M \geq 5$ ($M_{th}=5$) from Jan. 1, 2014 up to, but not including, the *M*8.8 mainshock. (**a**) ΔAIC versus candidate change point time $T_c$=0–4,234 days (days since Jan. 1, 2014; i.e., Jan. 1, 2014–Aug. 5, 2025). The blue band $T_c$=4,077–4,218 days (Mar. 1, 2025–Jul. 20, 2025) denotes the 68% confidence interval for the change point. Stable evaluation of $AIC_{2stage}$ across all $T_c$ was difficult, so we fixed α=1.27 (from Fig. 6), independent of $T_c$, and optimized (μ, $K_0$, α, *c*) separately before and after each $T_c$ to obtain $AIC_1$ and $AIC_2$, with $AIC_{2stage}=AIC_1+AIC_2$. The single-stage $AIC_{single}$ was computed under the same constraint (α=1.27) by optimizing (μ, $K_0$, α, *c*) over the full interval. Subtraction of $AIC_{2stage}$ from $AIC_{single}$ gives ΔAIC (=$AIC_{single}$-$AIC_{2stage}$). (**b**) Same as (**a**), but zoomed for $T_c \geq 4{,}000$ days (since Dec 12, 2024). The red circle marks $T_c$=4,218 days (Jul. 20, 2025); the corresponding ETAS fit is shown in Fig. 10.

We again observe ΔAIC≥2$q$, favoring the two-stage model (Fig. 9). The 68% confidence interval for the change point (blue band) is $T_c$=4,077–4,218 days since Jan. 1, 2014—Mar. 1, 2025 to Jul. 20, 2025. The maximum ΔAIC occurs at $T_c$=4,218 days (Jul. 20, 2025), indicating that the most probable onset of activation coincides with the $M$7.4 foreshock.

Using $T_c$=4,218 days, we fitted a single-stage ETAS model to events from Jan. 1, 2014 up to, but not including, the $M$8.8 mainshock (Fig. 10). For this fit, we set $S$=0.01 days, $T$=$T_c$=4,218 days (Jul. 20, 2025), and $T_{end}$=4,226.49 days (the date of the last $M$≥5 event before $M$8.8). During the prediction interval (after $T$), the observed rate (black) exceeds the extrapolated ETAS rate (red), consistent with seismic activation.

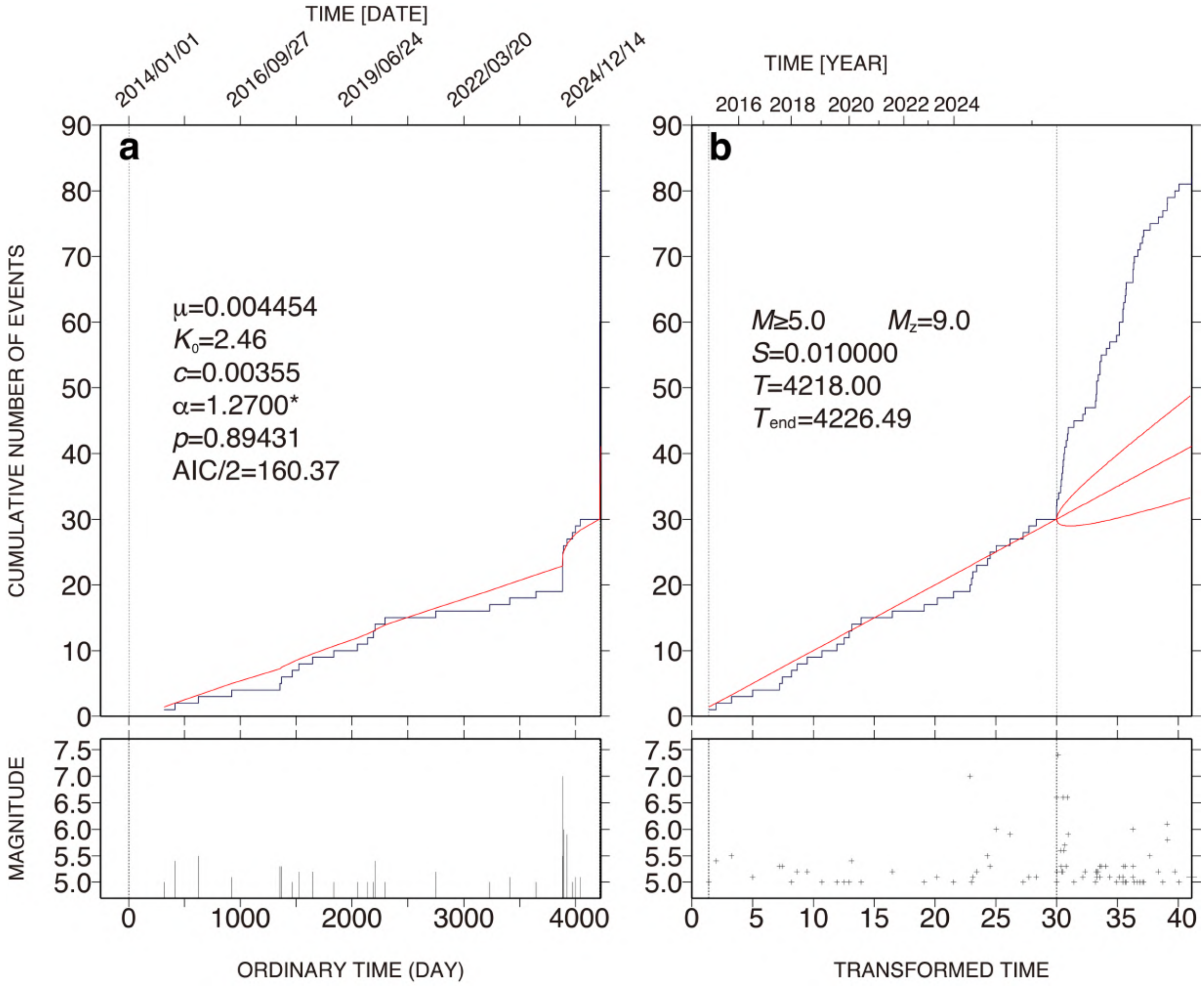

**Figure 10.** ETAS modeling for seismic activation (corresponding to the red-circled $T_c$ in Fig. 9). (**a**) Cumulative counts (black) and ETAS fit (red) in ordinary time for $M \geq 5$ ($M_{th}$=5) events within a 100-km radius of the *M*8.8 epicenter. Vertical dashed line marks *S*=0.01 days (Jan. 1, 2014 00:14:25) and *T*=4,218 days (Jul. 20, 2025). The parameter α is fixed at 1.27. (**b**) Same as (**a**), but in transformed time. The interval after *T*=4,218 days (Jul. 20, 2025 00:00:00) is the prediction period ($T$-$T_{end}$), where $T_{end}$=4,226.49 days (Jul. 28, 2025 11:45:04, the last event before the *M*8.8 earthquake). The observed curve exceeds the ETAS extrapolation, consistent with seismic activation.

## 5 Discussion and Conclusions

The 2025 *M*8.8 Kamchatka earthquake offers a unique opportunity to examine the temporal evolution of seismicity preceding a great subduction event. Using ETAS modeling and change-point analysis (Ogata & Tsuruoka, 2016; Kumazawa et al., 2010, 2019; Nanjo et al., 2023; 2025a,b; 2026), we identified two distinct phases within a 100~150-km radius of the mainshock epicenter: a long-term period of seismic quiescence that began around mid-2003 and persisted for roughly two decades (Figs. 4–6), followed by a short-term activation that commenced with the *M*7.4 earthquake on 20 Jul. 2025 (Figs. 4, 9, and 10). This sequence—from prolonged quiescence to abrupt activation—mirrors patterns documented for several large earthquakes, including the 1997 *M*7.8 Kronotsky, Kamchatka event (Sobolev, 2011, 2020), the 2004 *M*9.0 Sumatra event (e.g., Mignan et al., 2006; Sobolev & Lyubushin, 2007; Imoto & Yamamoto, 2008; Katsumata, 2015; Sobolev 2020), the 2006 *M*8.3 Simushirskoe, Kuril event (Sobolev, 2011, 2020; Katsumata & Zhuang, 2020), the 2010 *M*6.4 Jiashian, Taiwan earthquake (Wen et al., 2016), and the 2011 Tohoku event (e.g., Katsumata, 2011, 2017; Ogata, 2011; Kato et al., 2012; Nanjo et al., 2012; Nagao et al., 2014). The 2025 Kamchatka case is distinguished by the clarity of its temporal transitions, the spatial confinement of activity, and the availability of continuous data spanning more than five decades.

The nearly 20-year quiescent interval revealed by the ETAS analysis is commonly interpreted as reflecting stress accumulation along a locked plate interface or evolving physical properties within the fault zone (Scholz, 2019; Mignan, 2012). Laboratory friction experiments and numerical simulations indicate that a progressive decrease in microseismicity under sustained loading can be induced by asperity healing/strengthening (Braun & Peyrard, 2018) and stress corrosion (Main & Meredith, 1991), and by strain hardening of the crust, in which the closure and locking of microcracks and the exclusion of pore fluids stiffen the crust as stress rises, shifting microcracks from unstable stick-slip toward stable sliding and thereby suppressing background seismicity (Rundle et al., 2022). Our results establish, from seismicity alone, a spatiotemporal fact: within the source region of the 2025 *M*8.8 earthquake, an approximately two-decade quiescence preceded a short-term activation and the mainshock, and this quiescence is confined to the region that nucleated the rupture, being absent from the high-slip asperity to the southwest (Fig. 7d). What physical process produced it is a separate question that seismicity data alone cannot resolve, and we discuss it only as a possibility. A multiyear reduction in seismicity of this kind may reflect enhanced locking, in which the nucleation patch couples more strongly and accumulates stress toward failure; alternatively, it may reflect aseismic (slow) slip, which can either advance nucleation as precursory slip that concentrates stress on the adjacent locked patch, or relieve stress if it releases strain aseismically, as suggested for the small coseismic slip near the epicenter. Because the ETAS-based quiescence is a phenomenological diagnostic—the misfit measures the anomaly but not its cause—these possibilities cannot be distinguished from the seismicity data alone, and we do not favor one over another here. Resolving the mechanism would require continuous geodetic observation near the source region to detect or exclude aseismic slip, which is beyond the scope of this study; importantly, any such geodetic study of the mechanism presupposes the very quiescence-to-activation sequence that our seismicity analysis has established. Whichever process operates, the central result of this study—that quiescence and activation occurred within the source region of the *M*8.8 earthquake—is unaffected.

The two phases may be mechanistically linked. If the quiescence reflects a locked patch approaching failure, the ensuing activation would mark that patch crossing its threshold; if instead aseismic slip contributed to the quiescence, its acceleration into precursory slip could drive the same activation. In either reading, the quiescence-to-activation sequence represents a single preparatory process rather than two unrelated episodes, and distinguishing these pathways requires the near-source geodetic constraints noted above.

The subsequent activation phase began about 10 days before the mainshock, coincident with the *M*7.4 event. The rapid increase in the observed rate relative to the ETAS-extrapolated rate (λ) suggests that the *M*7.4 shock may have locally increased Coulomb stress (e.g., Lin & Stein, 2024; Toda et al., 2025) on that same patch, which subsequently ruptured as the *M*8.8 mainshock. Similar short-term accelerations prior to large earthquakes have been attributed to dynamic triggering, fluid-pressure transients, or aseismic-slip migration (e.g., Kato et al., 2012; Uchida & Matsuzawa, 2013; Peng & Lei, 2025; Zi et al., 2025). In Kamchatka, the sharp transition from long-term quiescence to abrupt activation may represent the final bridge between the interseismic and coseismic phases of the earthquake cycle.

A noteworthy aspect is the separation of temporal scales between the two phases. Multi-decadal quiescence reflects slow, large-scale stress accumulation, whereas the days-long activation reflects rapid, small-scale perturbations that may ultimately trigger rupture. This hierarchical behavior implies multiple interacting timescales—from decades-long loading to transients operating over days or hours. Recognizing and quantifying both regimes are essential for improving time-dependent earthquake-probability models. Integrating long-term quiescence as a proxy for stress accumulation and short-term activation as a precursory acceleration could yield a richer probabilistic framework for assessing the imminence of large earthquakes.

Sobolev (2011, 2020), using the RTL model, found that both the *M*7.8 Kamchatka earthquake of Dec. 5, 1997 (Bourgeois & Pinegina, 2018) and the *M*8.3 Simushirskoe earthquake of Nov. 15, 2006 (Katsumata & Zhuang, 2020) along the Kuril Arc (Fig. 1b,c) exhibited a temporal sequence from quiescence to activation. Prior to the 1997 event, quiescence developed in 1994–1995, followed by activation in 1995–1996, then a return to background levels. Before the 2006 event, quiescence occurred in 2001–2003, followed by activation in 2003–2005, again returning to the background. The 1997 event ruptured a region north of the 2025 mainshock, whereas the 2006 event occurred to its south (Fig. 1b,c).

The general sequence of quiescence followed by activation in these two cases is consistent with what we identified for 2025, suggesting that such quiescence-to-activation transitions may represent a recurring preparatory process for megathrust earthquakes in the Kamchatka–Kuril subduction system. Because this inference rests on only three events, which occupy adjacent segments and may not be fully independent, we regard it as a hypothesis to be tested rather than an established pattern.

Two differences are notable. First, the activation zone preceding the 2006 event was slightly offset from the eventual hypocenter (Sobolev, 2020), while in the 1997 and 2025 cases, the hypocenters overlapped with both the quiescent and activation areas. Second, for 1997 and 2006, there was a delay between the end of activation and the mainshock (Sobolev, 2020), whereas no such delay is evident in 2025.

Our findings underscore the value of high-resolution, long-duration seismic monitoring. Applying ETAS to multi-decadal catalogs enables quantitative detection of subtle changes in the seismicity rate (λ). The change point centered on Jun. 23, 2003 (68% confidence interval; Fig. 5b), together with the statistically significant deviation of the observed rate from the ETAS-extrapolated rate (approximately 95% confidence; Fig. 6), indicates that quiescence is not merely a visual impression but a quantifiable anomaly. Similarly, the change point (Jul. 20, 2025 within 68% confidence interval; Fig. 9), together with the statistically significant deviation (above 95% confidence; Fig. 10), indicates that activation is a quantifiable anomaly. Such diagnostics could become core components of real-time forecasting when coupled with Bayesian or machine-learning frameworks that assimilate continuous data streams.

More broadly, the Kamchatka 2025 sequence reinforces that the earthquake cycle evolves dynamically rather than stationarily. Long-term quiescence likely corresponds to strain/stress accumulation, whereas short-term activation reflects the final destabilization of the fault system. Monitoring transitions between these states may improve probabilistic assessments of a large-event likelihood, especially in the Kamchatka-Kuril subduction zone, and support more adaptive, transparent hazard communication.

A similar decadal quiescence—without an ensuing activation phase or a large earthquake to date—has recently been reported off eastern Hokkaido, Japan (Fig. 1a), at the southern part of the Kamchatka–Kuril subduction system (Matsu'ura, 2018; Matsu'ura et al., 2024; Nanjo et al., 2026). In that region, several independent observations have been documented in and around the quiescence area, including low and decreasing Gutenberg–Richter *b*-values, a seismic gap, strong plate coupling, and slow earthquake activity avoiding megaquakes' rupture zones (Nanjo et al., 2026). Notably, the quiescent region substantially overlaps the area assessed as having an elevated likelihood of recurrence of a 17th-century-type Hokkaido *M*9-class earthquake (Headquarters for Earthquake Research Promotion, 2017). Taken together with the three Kamchatka–Kuril cases discussed above (*M*7.8 in 1997, *M*8.3 in 2006, and *M*8.8 in 2025), these observations raise the possibility that the currently observed quiescence off Hokkaido could eventually transition to an activation phase prior to a future large earthquake, although this should not be interpreted deterministically. Accordingly, continued, multidisciplinary monitoring along the plate interface off the Pacific coast of Japan remains essential to improve situational awareness and support transparent, time-dependent hazard assessment.

Limitations must be acknowledged. Foremost, this analysis is retrospective: the source region, study period, and interpretation were all established after the 2025 mainshock, so it documents the seismicity evolution of a single, exceptionally well-recorded megathrust event and does not constitute a validated prediction; assessing the predictive value of a quiescence-to-activation sequence would require prospective, region-wide testing across the Kamchatka–Kuril subduction system, which is beyond the scope of this case study.

Precursory quiescence is, moreover, neither universal nor exclusive: not every large earthquake is preceded by it, and not every quiescence is followed by one. The present cases are therefore best interpreted as instances of a documented but imperfect precursor, and a systematic survey of large earthquakes across the Kamchatka–Kuril system—including events that lack precursory quiescence—together with the prospective testing noted above, is required to quantify how often the sequence occurs with and without a subsequent mainshock.

Our region-based analysis samples the spatial variation of quiescence at a few representative locations; a fully continuous map, obtained by evaluating the change-point statistic within a fixed-radius circle moved across a fine (e.g., 0.1°) grid, would refine this spatial picture and is left for future work.

Two further methodological caveats apply. First, the ETAS framework assumes (piecewise) stationary parameters, whereas real fault systems may exhibit slow temporal evolution in $\theta$ due to external forcing (e.g., fluid flux or viscoelastic relaxation). Second, although the temporal coincidence between the activation onset and the *M*7.4 event is striking, causality remains to be tested with stress-transfer (e.g., Lin and Stein, 2024; Toda et al., 2025) and geodetic modeling. Future work should integrate high-resolution moment tensors and continuous geodetic observations to better constrain the spatiotemporal evolution of stress and coupling prior to the mainshock.

**Acknowledgments.** We thank four reviewers for their constructive comments. Research by K.Z.N. was partially supported by Japan's Ministry of Education, Culture, Sports, Science and Technology (MEXT) under the Second Earthquake and Volcano Hazards Observation and Research Program (Earthquake and Volcano Hazard Reduction Research), and under the STAR-E (Seismology TowArd Research innovation with data of Earthquake) Program, Grant Number JPJ010217; the Takahashi Industrial and Economic Research Foundation (FY2025 Research Grant); and the SECOM Science and Technology Foundation (FY2023 Special Research Grant). Research by J.B.R. was supported in part by the Southern California Earthquake Center (SCEC) under grant no. SCON-00007927 to UC Davis, and by the John LaBrecque Fund, a generous gift to the University of California, Davis.

**Open Research.** Earthquake catalogs were downloaded from the USGS ANSS database (USGS, 2017) at https://earthquake.usgs.gov/earthquakes/search/. Figures 1, 5e, and 7e were prepared with the Generic Mapping Tools (GMT; https://www.generic-mapping-tools.org) (Wessel et al.,2013), using the digital plate-boundary model of Bird (2003) from https://www.earthbyte.org/category/resources/data-models/plate-boundaries/ and the ETOPO Global Relief Model from https://www.ncei.noaa.gov/products/etopo-global-relief-model

(NOAA National Centers for Environmental Information, 2022). Figure 3 was genaretd with the ZMAP package (Wiemer, 2001). Figures 4-10 were generated with the XETAS GUI package (Ogata & Tsuruoka, 2016). Softwares to reproduce Figures 2, 3, 5, 7, and 9 are archived on figshare at https://doi.org/10.6084/m9.figshare.33411553, https://doi.org/10.6084/m9.figshare.33411577, and https://doi.org/10.6084/m9.figshare.33411583 (Nanjo, 2026a,b,c).

**Conflict of Interest Disclosure.** The authors declare there are no conflicts of interest for this manuscript.

**CRediT.** Conceptualization (K.Z.N., J.T., I.T.B, J.B.R.), Data curation (K.Z.N.), Formal analysis (K.Z.N.), Funding acquisition (K.Z.N., J.B.R.), Investigation (K.Z.N., J.T.), Methodology (K.Z.N., J.T., I.T.B), Project administration (K.Z.N.), Resources (K.Z.N.), Software (K.Z.N.), Supervision (K.Z.N., J.B.R.), Validation (K.Z.N.), Visualization (K.Z.N.), Writing – original draft (K.Z.N.), Writing – review & editing (K.Z.N., J.T., I.T.B, J.B.R.).